# Thermal Desorption of $H_2O$-Ice: From Nanoscale Films to the Bulk

A. Rosu-Finsen,* B. Chikani, and C. G. Salzmann

Department of Chemistry, University College London, 20 Gordon Street, WC1H 0AJ, United Kingdom

*corresponding author: a.rosu-finsen@ucl.ac.uk

## Abstract

Desorption of $H_2O$ films ranging from 53 nanometres to 101 micrometre thicknesses have been investigated using a quartz-crystal microbalance (QCM) and temperature-programmed desorption. Three desorption stages are observed belonging to amorphous solid water (ASW), stacking disordered ice I (ice I*sd*), and hexagonal ice I (ice I*h*). The desorption of ASW is only visible for the ≥10 μm films and is separated from the ice I desorption by 10–15 K and has an associated desorption energy of ~64 kJ mol$^{-1}$. The desorption energy of the 53 nm film was found to be near 50 kJ mol$^{-1}$ as also noted in the literature, but with increasing film thickness the desorption energy of ice I rises until reaching a plateau around 65–70 kJ mol$^{-1}$. The reason for the increased desorption energy is suggested to be due to molecules unable to desorb due to the thick covering layer of $H_2O$ and possibly re-adsorption. Before complete desorption of ice I which occurs around 220 K for the 100 μm film, a two-stage ice I desorption is observed with the QCM for the 10 and 20 μm films near 200 K. This event corresponds to the desorption of ice I*sd* as corroborated by X-ray diffraction patterns collected upon heating from 92–260 K at ambient pressure. Cubic ice is not observed as is commonly stated in the literature as resulting from the crystallisation of ASW. Therefore, ice I*sd* is the correct terminology for the initial crystallisation product of ASW.

## 1 Introduction

Amorphous solid water (ASW) was discovered in the 1930s by vapour deposition onto a cryogenically cooled copper substrate (Burton and Oliver 1935a, 1935b). ASW is one type of low-density amorphous (LDA) ice, but LDA can be made *via* different methods such as decompression or heating of high-density ice (HDA) (Mishima, Calvert, and Whalley 1984, 1985). The notion of a LDA-I and LDA-II was brought forward when differently treated HDA ices were transformed to LDA and analysed through diffraction (Winkel et al. 2009), but vibrational spectroscopic measurements were not able to distinguish the difference between such LDA types (Shephard, Evans, and Salzmann 2013). LDA can also be made by rapidly cooling liquid water forming hyperquenched glassy water (HQW) (Brüggeller and Mayer 1980; Mayer and Brüggeller 1982), or when ice VIII is heated at ambient pressure (Klug et al. 1989). Interestingly, spectroscopy can distinguish between LDA from ice VIII and other LDA ices (Shephard et al. 2016).

Depending on the preparation conditions such as molecular beam or background dosing, deposition rate and angle, the morphology of ASW is significantly affected (Mayer and Pletzer 1984; Stevenson et al. 1999; Kimmel et al. 2001; Dohnálek et al. 2002). Another important parameter to control is the temperature where porous or compact ASW form depending on the substrate temperature (Narten, Venkatesh, and Rice 1976; Jenniskens and Blake 1994). It is important to note that the high density vapour-deposited polyamorph is not similar to the compressed high density amorphous ice (Mishima, Calvert, and Whalley 1984). Regardless of the morphology and density of ASW, this polyamorph is the most abundant solid state species in the interstellar medium (Williams, Fraser, and McCoustra 2002; Tielens 2013) where reactive accretion onto dust grains in dark molecular clouds is the main formation pathway (Linnartz, Ioppolo, and Fedoseev 2015). Ice has also been detected on a range of extra-terrestrial objects such as comets, satellites, and planets (De Sanctis et al. 2015; Grundy et al. 2016; Filacchione et al. 2016).

As highlighted recently (Potapov, Jäger, and Henning 2020), based on $H_2O$ column densities of $H_2O$ (Gibb et al. 2004; Boogert et al. 2008; Boogert et al. 2011) it was estimated that 300–12,000 monolayers (ML) can be present on interstellar dust grains. With the density of LDA being 0.93 g cm$^{-3}$ (Mishima, Calvert, and Whalley 1985), these layers can be estimated as being about 90–360 nm thick. Interstellar dust grains range in size from the nanometre to micrometre level (Mathis, Rumpl, and Nordsieck 1977) and an ice coating of similar thickness is reasonable to consider. Icy, sticky dust grains coagulate, forming larger particles and ultimately comets and asteroids (Wada et al. 2009; Gundlach and Blum 2015). As the size of grains, comets, and asteroids grows, so too can the thickness of ice. Such agglomerated comets and asteroids may be highly porous where a range of high-energy binding sites can be present. Recent experiments of $H_2O$ physisorbed onto highly porous silicate grain material showed a drastic delay in desorption up to 200 K (Potapov et al. 2021). Likewise, $H_2O$ mixed with $C_{60}$-fullerenes (Halukeerthi et al. 2020) and fullerene-like carbon grains (Potapov, Jäger, and Henning 2018) have shown suppression or delays in desorption. The change in $H_2O$ desorption behaviour is believed to occur as the percolation threshold of the guest species is approached (Halukeerthi et al. 2020), and similar concentration-dependent behaviour has been noted in the photodesorption of $H_2O$-antracoronene mixtures (Korsmeyer et al. 2022).

Sub-monolayer coverages of $H_2O$ have been shown to produce clusters (Carrasco, Hodgson, and Michaelides 2012; Rosu-Finsen et al. 2016; Marchione et al. 2019) explaining why $H_2O$ desorption commonly occurs through zero-order kinetics (Fraser et al. 2001; Collings et al. 2001; Smith, Matthiesen, and Kay 2014; Collings et al. 2015). ASW contains internal pores (Mitterdorfer et al. 2014; Hill et al. 2016), and recent experiments have proposed that the collapse of such pores near 120-130 K constitute a new desorption pathway for trapped guest species only seen in thick ASW films (Talewar 2019). Common for both thin and thick ASW films with guest species, is the expulsion of guests during crystallisation (Collings et al. 2003; Collings et al. 2004; Talewar 2019) which has been termed the molecular volcano (Smith et al. 1997; May, Smith, and Kay 2012). Since pure $H_2O$ desorption has been shown to peak as high as 180 K (depending on heating rate and film thickness), the ASW sublimation feature during crystallisation usually overlaps with the overall $H_2O$ desorption profile (Fraser et al. 2001; Collings et al. 2001; Smith, Matthiesen, and Kay 2014; Collings et al. 2015).

When considering interstellar $H_2O$-ice, the conditions afforded by an ultrahigh vacuum system are ideal. However, as dust grains agglomerate into larger structures, the ice thicknesses will increase, and ultrahigh vacuum systems do not cope well with desorption of large quantities of adsorbates. In this work we combine a quartz crystal microbalance (QCM) with temperature-programmed desorption (TPD) to explore the sublimation behaviour of $H_2O$ films spanning from 53 nm to 101 μm in a high-vacuum chamber. With such bulk materials we also study the transition of ASW to hexagonal/cubic stacking disordered ice I (ice I*sd*), as well as the final formation of the stable hexagonal ice I (ice I*h*) through ambient pressure X-ray diffraction.

## 2 Experimental

The experimental equipment and analytical methods used in this work have been detailed previously (Talewar 2019; Halukeerthi et al. 2020), however, a brief description will follow here. A diffusion pump (Diffstak 63/150M, BOC Edwards) backed by a rotary pump (model 12, BOC Edwards) is used to routinely achieve a base pressure of $<2\times10^{-7}$ mbar in a high-vacuum chamber (Kurt J. Lesker Ltc., 12×12×24 inch internal dimension) as measured by a dual combination cold cathode/Pirani pressure gauge (PenningVac PTR 90, Oerlikon Leybold Vacuum). An inlet tube covered with a metal mesh attached to a high-precision needle valve (EV 016 DOS AB, Oerlikon Leybold Vacuum) controlling the amount of ultrapure $H_2O$ (MiliQ, Milipore) vapour allowed into the chamber at a constant pressure. Before use, the $H_2O$ was further purified by several freeze-pump-thaw cycles. A liquid nitrogen cooled 8-inch diameter copper deposition substrate place in-line with the inlet tube routinely reaches a base temperature of 84 K as measured by a K-type thermocouple connected to a purpose-made temperature read-out unit containing an Adafruit Feather 32u4 Basic Proto microcontroller and an Adafruit MAX31856 Universal Thermocouple Amplifier. When desired, the deposition plate can be removed from the vacuum chamber once an icy film has been deposited for further *ex-situ* analysis at ambient pressure. Attached to the deposition plate is a quartz crystal microbalance (QCM) fitted with a gold-plated AT-cut 6 MHz planoconvex quartz crystal placed inside an Allectra 710-SH sensor. The QCM sensor was connected to a reflection bridge and a 0.5–60 MHz N2PK vector network analyser. The QCM was used to monitor the film thickness in real time and deposition ceased once a desired film thickness was achieved. Temperature-programmed desorption (TPD) was recorded by a residual gas analyser mass spectrometer (RGA-MS, Hiden Analytical, HALO 201) monitoring the partial pressures in the 10–35 *m/z* range.

An experiment involving a recovered ASW sample was also conducted. Following ASW deposition, the ice was annealed at 125 K to close the pores in ASW and prevent $N_2$ hydrate formation when extracting the sample with liquid nitrogen at ambient pressure (Mayer and Hallbrucker 1989). The ASW was transferred into an X-ray sample holder with Kapton windows under liquid nitrogen conditions which was then mounted on a Stoe Stadi P X-ray diffractometer with Cu $K_\alpha 1$ radiation at 40 kV, 30 mA and monochromated by a Ge 111 crystal. The temperature of the ASW was controlled by an Oxford Instruments CryoJetHT. X-ray diffraction patterns were collected at a base temperature of 92 K and then in 10 K steps from 100 K until the melt by a Mythen 1K detector.

**3 Results and Discussions**

3.1 Quartz crystal microbalance

Figure 1(a) shows the decrease in frequency as ASW is deposited onto the QCM crystal. Since the deposition rate is kept constant, the change in frequency ($\Delta f$) is linear as seen in Figure 1(a) where the change in mass ($\Delta m$) of $H_2O$ deposited on the QCM crystal can be calculated from the Sauerbrey equation, eq. 1,

$$\Delta f = -\frac{2f_0^2}{A\sqrt{\rho_q \mu_q}} \times \Delta m \qquad \text{eq. 1}$$

where $f_0$ is the resonant frequency of the crystal, $A$ is the active crystal area, and $\rho_q$ and $\mu_q$ are the density and shear modulus of quartz, respectively. With increasing film thickness, the frequency change becomes larger reaching a point whereafter the Sauerbrey equation only becomes an approximation. This is due to the large shift in the frequency itself as well as the possible influence of viscoelastic effects which become important when the film thickness is a sizeable fraction of the wavelength of the shear sound. With the limitations of eq.1 in mind, we continue to calculate the film thicknesses as shown in Table 1 based on the mass determined from eq. 1 and the density of ASW (0.93 g cm$^{-3}$) (Mishima, Calvert, and Whalley 1985).

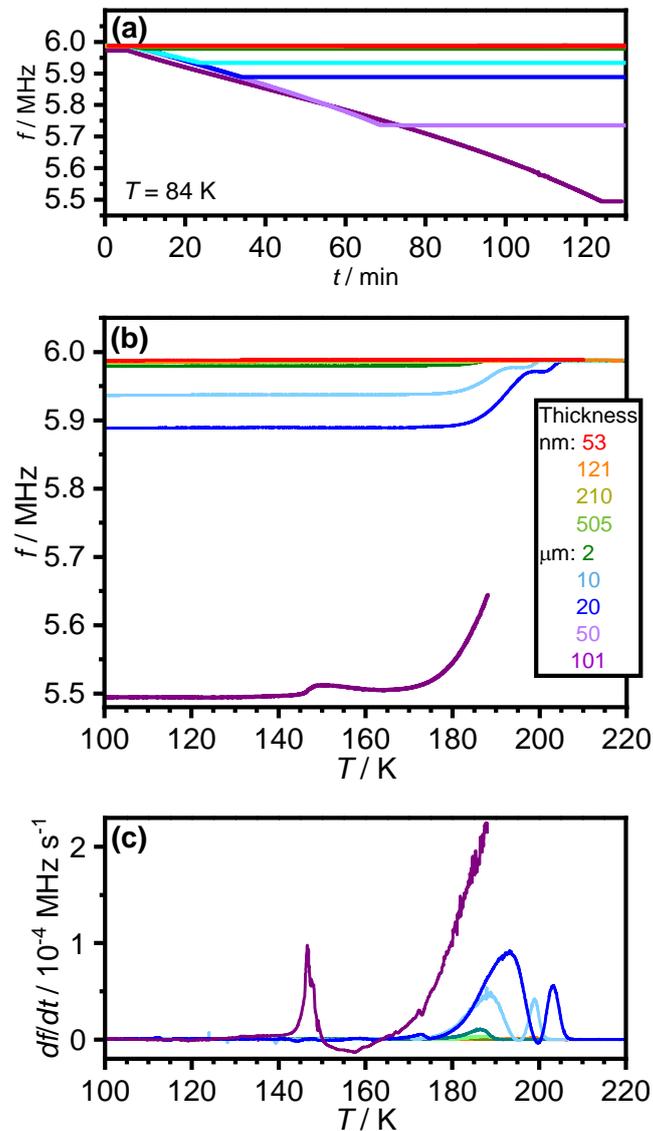

*Figure 1: (a) Decrease in frequency upon ASW deposition at 84 K. (b) Changes in frequency upon heating H$_2$O films of increasing thicknesses. (c) Differentiated and 25-point adjacent-averaged smoothed QCM signal with respect to time but plotted against temperature. Loss of mass is noted at 150 K for the 101 μm film, and full ice I desorption through a two-stage process starting near 170 K and 200 K depending on film thickness. The colour legend in panel (b) is valid for all panels.*

*Table 1: Overall frequency changes upon deposition which are used to determine the mass of H₂O deposited onto the entire plate. The film thicknesses and numbers of ML are based on the mass and density of ASW.*

| $\Delta f$ / Hz | $\Delta m$ / mg | Film Thickness | H$_2$O monolayers / ML |
|---|---|---|---|
| -261 | 2 | 53 nm | 178 |
| -593 | 4 | 121 nm | 405 |
| -1,021 | 6 | 209 nm | 698 |
| -2,464 | 15 | 505 nm | 1,684 |
| -11,012 | 51 | 2 μm | 7,526 |
| -50,127 | 309 | 10 μm | 34,258 |
| -98,193 | 606 | 20 μm | 67,108 |
| -250,414 | 1,545 | 51 μm | 171,140 |
| -494,690 | 3,051 | 101 μm | 338,084 |

Following deposition, the amorphous ice was heated with the QCM monitoring the desorption as shown in Figure 1(b) and Figure 1(c) of the adjacent-averaged smoothed derivative of the QCM signal (*df/dt*). Near 150 K, ASW desorbs as the free energy of ASW is greater than that of its crystalline form (Kouchi 1987; Sack and Baragiola 1993; Smith et al. 2012; Nachbar, Duft, and Leisner 2018) leading to a greater vapour pressure of ASW. As such, crystallisation, in essence, prevents the full desorption of ASW resulting in the desorption peak in panel (c) for the 100 μm film. As this technique concerns frequency changes due to variations in mass, the loss of H$_2$O can be quantified at ~4% for the 100 μm film. It is peculiar to note that only the 100 μm film exhibits this desorption event which may indicate that a threshold has been reached between 20 and 100 μm. As H$_2$O is known to have poor thermal conductivity (Kouchi et al. 1992), it is possible that the exothermic nature of crystallisation leads to local heating predominately at the surface of the thickest film thereby yielding an observable desorption feature near 150 K. Unfortunately, the stability of the QCM signal reduces drastically with large deviations from the fundamental frequency (in this case $\Delta f$ is as large as ~0.5 MHz) meaning that the software easily loses the QCM signal. This is the reason for the loss of desorption data for the 50 μm film, as well as the loss of signal for the 100 μm film above 190 K. Following crystallisation, a slight decrease in frequency can be seen in Figure 1(b) and a drop below the baseline in panel (c). This frequency decrease has previously been observed (Talewar 2019) and discussed as possibly arising due to the release of bending stresses and changes in the viscoelastic properties of thick films of $H_2O$.

A two-stage desorption event for ice I is clearly seen in panel (c) for the 10 and 20 μm films which indicates the possible transition from one solid-state species to another with different vapour pressures and coincide with the ice I$sd$ to ice I$h$ transition (Kuhs, Bliss, and Finney 1987; Malkin et al. 2015). Ice I$sd$ is metastable with respect to ice I$h$ and has a higher free energy and therefore a higher vapour pressure meaning ice I$sd$ desorbs at a lower temperature. As such, it is interesting to note that only the ≥10 μm films exhibit conversion of ice I$sd$ to ice I$h$ followed by ice I$h$ desorption after which the QCM signal reaches 6 MHz indicating that the crystal and deposition plate are once again bare. In the case of the thinner films, complete desorption is achieved before the ice I$sd$ to ice I$h$ phase transition.

3.2 X-ray diffraction of vapour-deposited $H_2O$

The ice I family is comprised of structures ranging from only hexagonal stacking, called ice I$h$, to the entirely cubic stacking in ice I (ice I$c$). Between these end-members is what has been termed ice I$sd$ (Kuhs et al. 2012; Malkin et al. 2015) along with a range of more complex and periodic stacking sequences such as the 4$H$, 6$H$, and 9$R$ polytypes (Salzmann and Murray 2020). Ice I$h$ is the thermodynamically stable phase of ice at ambient pressure, hence all phases of ice eventually transform to this with increasing temperature. In the literature, the erroneous description of ASW transforming to ice I$c$ at 150 K followed by an ice I$c$→ice I$h$ transition at higher temperatures is commonplace. Ice I$c$ has only recently been prepared in its pure form following complex preparation procedures (del Rosso et al. 2020; Komatsu et al. 2020).

To illustrate the transformation of ASW to ice I$sd$, Figure 2 shows the X-ray diffraction data (in black) upon heating ASW at ambient pressure from 92 to 260 K.

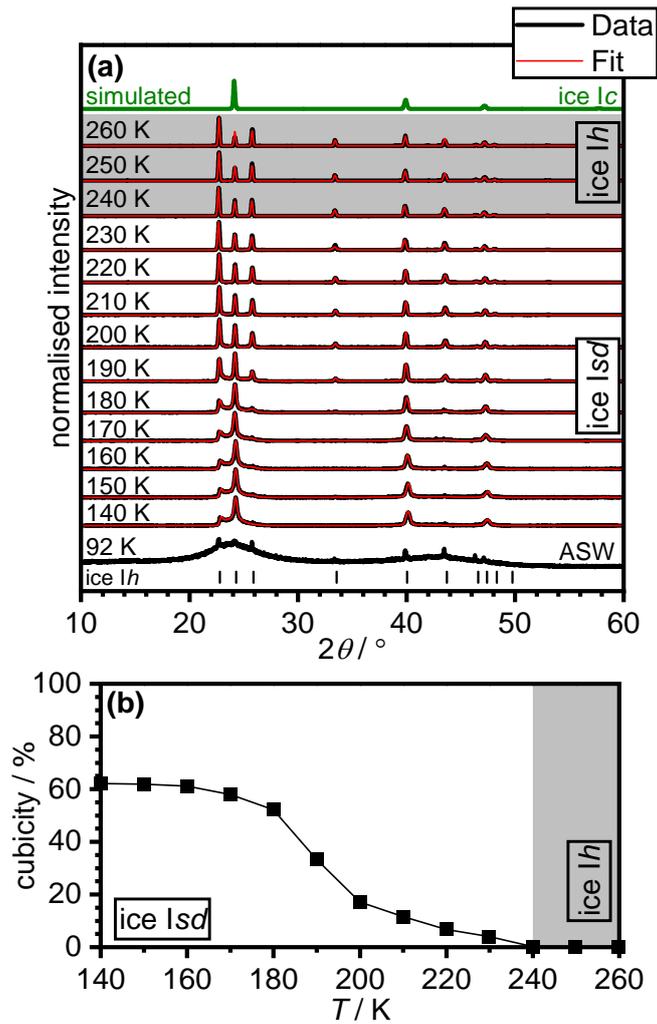

*Figure 2. (a) Diffraction patterns collected upon heating ASW at ambient pressure (black). Crystallisation to ice Isd is seen by the sharpening of peaks and formation of ice Ih only occurs as the diffuse scattering disappears in the 24-27° range. The red lines indicate MCDIFFaX fits. (b) Changes in cubicity of ice Isd with the formation of ice Ih near 240 K.*

The experimental data in Figure 2(a) was obtained from a 100 μm ASW film prepared in the vacuum chamber and extracted to ambient pressure at 77 K. The broad features at 92 K near 24° and 44° are indicative of LDA (Burton and Oliver 1935b), and therefore ASW, meaning the ice prepared in the vacuum chamber was successfully extracted and transferred to the diffractometer. Upon heating, as shown previously (Talewar 2019; Halukeerthi et al. 2020), ASW crystallises to ice I*sd*. If a perfectly crystalline structure was formed, sharper and baseline separated Bragg peaks would be expected. However, this is not observed in Figure 2(a) until >200 K where ice I*sd* begins to transform to ice I*h*. For ice I*h*, the baseline separated "trident" between 23–26° is observed. Ice I*c* only displays the 24° Bragg peak (del Rosso et al. 2020; Komatsu et al. 2020; Salzmann and Murray 2020) as shown by the simulated ice I*c* data in green in panel (a). From this, Figure 2(a) clearly shows experimental diffraction patterns that are not pure ice I$c$, but where the diffuse scattering between the trident peaks indicates stacking disorder. To further analyse the presence of hexagonal and cubic stacking sequences, the DIFFaX software with an added Monte Carlo algorithm (MCDIFFaX) (Salzmann, Murray, and Shephard 2015) was used to fit the X-ray diffraction data in a least-squares environment. To optimise the fit and reach a $\chi^2$ convergence, the software searches for the best values for the lattice constants, peak profile parameters, and zero-shift as well as the optimal stacking probabilities. The red lines in Figure 2(a) are the resultant fits, and from this, the stacking probabilities are obtained including the percentage of cubic stacking events called cubicity shown as a function of temperature in Figure 2(b). As can be seen, the first pattern of ice I$sd$ only yields a cubicity of 60% whereafter the cubicity decreases to zero near 240 K as pure ice I$h$ forms. From the diffraction data, no other transitions or structural changes are observed indicating that the "wiggle" in the QCM data in Figure 1(b) and the two-stage desorption event in Figure 1(c) must arise from the ice I$sd$ to ice I$h$ transition. The minor differences in transition temperatures between the different analytical techniques are due to different heating rates and pressure conditions. However, the overall trend is clear: ASW crystallises to a stacking disordered ice I, but not ice I$c$.

3.3 Temperature-programmed desorption of H$_2$O
The thinnest film studied in this work is 53 nm thick exhibiting a single desorption feature peaking at 178 K as seen in Figure 3(a) which is similar to previous desorption data published in the literature (Fraser et al. 2001; Collings et al. 2015; Potapov, Jäger, and Henning 2018; Salter et al. 2021).

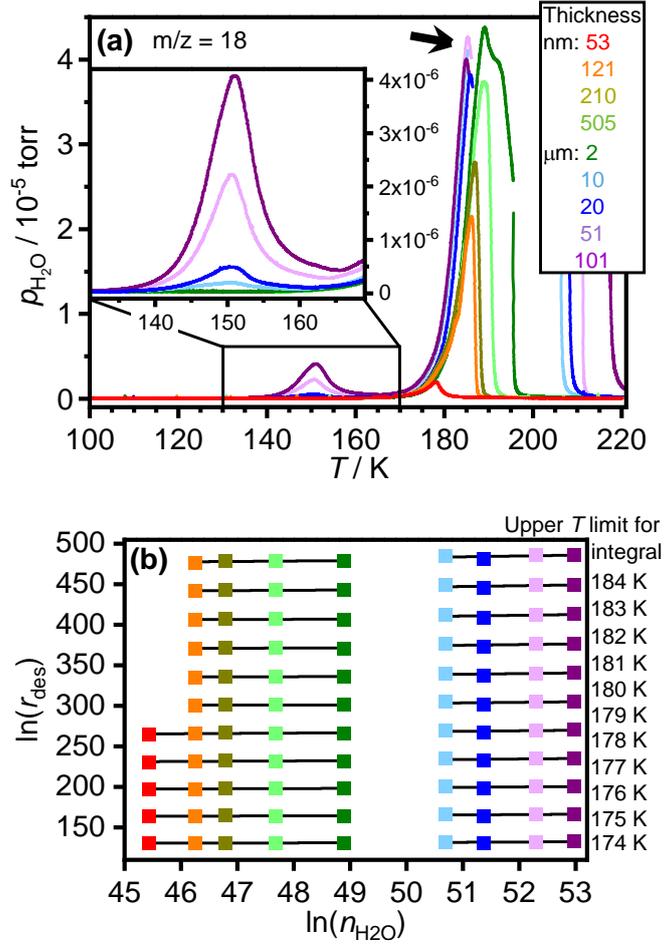

*Figure 3: (a) Temperature-programmed desorption traces of $H_2O$ (m/z = 18) at increasing film thicknesses ranging from 53 nm to 101 μm. Initially, the traces feature only one desorption event which are split into two separate events at increasing film thicknesses. The insert in (a) highlights the desorption of ASW of the ≥10 μm films. The black arrow highlights the drop in partial pressure during ice Isd desorption. (b) Leading-edge analysis indicates zero-order desorption kinetics for all film thicknesses. The colour coding across the two panels indicated $H_2O$ film thickness.*

With increasing film thickness, the desorption peak is expected to shift to higher temperatures for adsorbates following zero-order desorption kinetics. This behaviour is seen in Figure 3(a) as the desorption peak shifts to ~190 K for the 500 nm film, where a single desorption event is observed for all film thicknesses up to 2 μm. At ≥10 μm, a separate ASW desorption feature peaking near 150 K appears which grows with increasing $H_2O$ thickness. This desorption feature has previously been observed, but commonly as a small shoulder on the low-temperature side of $H_2O$ desorption traces (Fraser et al. 2001; Smith et al. 2011; Collings et al. 2015; Potapov, Jäger, and Henning 2018; Salter et al. 2021). While there is a slight separation between the 2 and 10 μm TPD traces, $H_2O$ desorption starting near 170 K display coincident leading edges on either side of the separation indicative of multilayer desorption following zero-order kinetics shown in Figure 3(b) and expected for $H_2O$ (Fraser et al. 2001; Collings et al. 2001; Smith, Matthiesen, and Kay 2014; Collings et al. 2015). Figure 3(b) is the result of a leading-edge analysis where the molecules desorbed at certain temperatures are integrated. The gradient of the individual integrals for each TPD trace at a certain temperature determines the kinetic desorption order which is 0.29±0.40 in this work. The reason for the shift in the desorption leading edge can be seen in the increased heating rate from 0.50 min$^{-1}$ for the 2 μm film to 0.73 K min$^{-1}$ for the 10 μm film. With increasing amounts of $H_2O$ desorbed from the deposition plate, the pressure in the chamber increases leading to a more effective heat exchange between the chamber walls and the cryostat. This continued bombardment of gas-phase molecules could explain the higher heating rates observed for the thicker samples. Following the complete desorption of the 100 μm film, the heating rate decreases to 0.32 K min$^{-1}$ lending credence to this idea.

The sharp shoulder near 195 K of the 2 μm TPD trace in Figure 3(a) sets it apart from the thinner film traces where single, clean peaks are displayed. The shoulder in the 2 μm TPD data could indicate the beginning of ice I$h$ formation upon heating, however, a desorption signal was not observed in Figure 1(b, c) which may be due to only a small amount of ice I$h$ being formed and desorbed. As the thickness increases, so does the amount of $H_2O$ desorbed and the chamber pressure rises to an extent too great for the safe operation of the mass spectrometer which was temporarily turned off. However, a drop in partial pressures of the ≥10 μm is noted near 185 K as highlighted by the black arrow in Figure 3(a). At this stage the pressure continues to rise due to $H_2O$ desorption where Figure 1(b, c) show that $H_2O$ is still present in the solid state. It is therefore likely that the drop in the $H_2O$ partial pressure near 185 K is due to the desorption of ice I$sd$ whereafter ice I$h$ is formed and begins desorption leading to the increased chamber pressure. Following ice I$h$ desorption of the ≥10 μm films, the chamber pressure decreases rapidly, and the spectrometer was turned on again to collect the trailing desorption edge highlighting that desorption of the 100 μm film terminates at 220 K as shown in Figure 3(a).

3.4 Kinetic simulations of $H_2O$ desorption

The kinetics related to $H_2O$ desorption were simulated with the Kinetiscope stochastic kinetics simulator (Kinetiscope) and compared to the experimental data as shown in Figure 4(a).

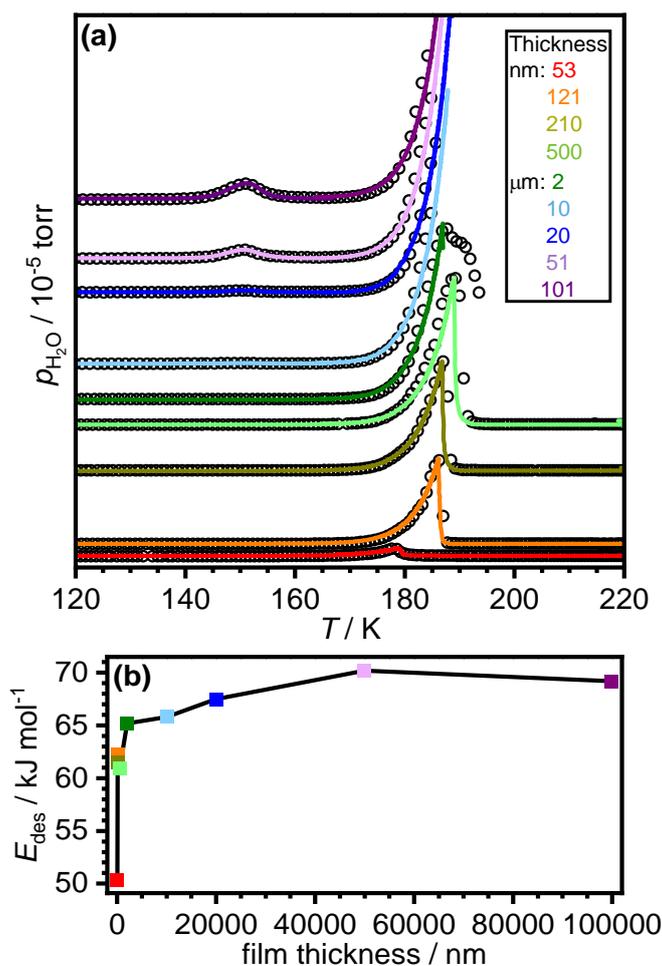

*Figure 4: (a) Simulated (solid lines) and experimental data (black open circles). (b) the desorption energy of $H_2O$ as a function of film thickness in nanometres. The colour coding in panel (a) indicates the film thickness.*

The reaction mechanisms in eq. 2-4 indicate the physical processes occurring during desorption where *(s)* and *(g)* relate to the physical state of $H_2O$ being solid or gas, respectively, and as such, were used with the Kinetiscope software.

$$ASW(s) \rightarrow H_2O(g) \qquad \text{eq. 2}$$

$$ice\ I(s) \rightarrow H_2O(g) \qquad \text{eq. 3}$$

$$H_2O(g) \rightarrow pump \qquad \text{eq. 4}$$

Eq. 2 relates to the desorption of ASW while eq. 3 is of ice I desorption and eq. 4 is the removal of *$H_2O(g)$* by the pumps. The initial concentration of *ASW(s)* and *ice I(s)* for the simulations in terms of molecules $cm^{-3}$ were based on the QCM measurements and extrapolated for the area of the entire deposition plate. The recorded data of the heating rate from each individual experiment was loaded directly into Kinetiscope. The desorption energy ($E_{des}$) and pre-exponential factor (*v*) were gathered from an inverse Arrhenius equation shown in eq. 5 and used as the initial parameters for the simulation

$$\ln(k) = \ln(v) \times \frac{-E_{des}}{RT} \qquad \text{eq. 5}$$

where *k* is the rate constant and *RT* is the product of the ideal gas constant and temperature. The obtained desorption energies from eq. 5 are shown in Table 2 alongside those extracted from Kinetiscope and heating rates during ice I desorption.

*Table 2: Experimental heating rate during desorption which vary with amount of deposited $H_2O$. An initial estimate of the desorption energy from an inverse Arrhenius analysis is in line with the extracted parameters gathered from the Kinetiscope software shown in the two last columns.*

| Film thickness | Heating Rate / K min$^{-1}$ | Inverse Arrhenius analysis / kJ mol$^{-1}$ | $E_{des}$ / kJ mol$^{-1}$ | $v_{des}$ / molecules cm$^{-2}$ s$^{-1}$ |
|---|---|---|---|---|
| 53 nm | 0.43 | 48 | 50.3 | 1×10$^{30}$ |
| 100 nm | 0.45 | 59 | 62.2 | 1×10$^{30}$ |
| 210 nm | 0.45 | 61 | 61.5 | 1×10$^{30}$ |
| 500 nm | 0.47 | 63 | 60.9 | 1×10$^{30}$ |
| 2 µm | 0.50 | 64 | 65.2 | 1×10$^{32}$ |
| 10 µm | 0.73 | 67 | 65.8 | 1×10$^{33}$ |
| 20 µm | 0.82 | 67 | 67.5 | 5×10$^{33}$ |
| 51 µm | 0.89 | 66 | 70.2 | 1×10$^{35}$ |
| 101 µm | 0.92 | 66 | 69.2 | 1×10$^{35}$ |

As mentioned previously, a combined peak related to ASW desorption and crystallisation commonly occurs during $H_2O$ desorption, however, the separate desorption events shown in Figure 3(a) are simulated in Figure 4(a). A best-fit with Kinetiscope leads to an ASW desorption energy of about 64.2±0.3 kJ mol$^{-1}$ and a pre-exponential factor of about 9×10$^{17\pm1}$ s$^{-1}$. This is higher than the 10$^{12}$ s$^{-1}$ typically assumed for physisorbed species, however, it is close to the 1.1×10$^{18}$ s$^{-1}$ pre-exponential factor for 1,000 ML ASW previously determined (Smith et al. 2011). In that study the desorption energy was determined to be about 55 kJ mol$^{-1}$ (Smith et al. 2011) which is lower than the desorption energy found in this work.

Simulations of the ice I desorption features were also conducted as shown in Figure 4(a). The desorption energy and pre-exponential factor of the 53 nm film as shown in Table 2 are in good agreement with the literature (Fraser et al. 2001; Collings et al. 2015; Potapov, Jäger, and Henning 2018; Salter et al. 2021). However, as the thickness increases, so too does the desorption energy. The desorption energies following the 2 μm film appear to form a plateau in the 64–67 kJ mol$^{-1}$ range. It is important to note that complete ice I desorption data of the >2 μm films could not be collected due to the high chamber pressure during H$_2$O desorption, however, the leading edges have been simulated to the best extent possible in Figure 4(a). The average desorption energy (65.3 kJ mol$^{-1}$) and pre-exponential factor ($2\times10^{34}$ molecules cm$^{-2}$ s$^{-1}$) being greater than the literature (commonly found to be near 50 kJ mol$^{-1}$ and $10^{28}$-$10^{30}$ molecules cm$^{-2}$ s$^{-1}$) (Fraser et al. 2001; Collings et al. 2015; Potapov, Jäger, and Henning 2018) will be discussed later.

**4 Astrophysical Implications**

Investigating the kinetics of physical processes such as desorption is of crucial importance for furthering our understanding of the astrophysical environment as recently discussed in a detailed review (Minissale et al. 2022). As H$_2$O is the most abundant solid in the interstellar medium, it has attracted a lot of attention and been thoroughly studied. However, the thick film/bulk H$_2$O desorption behaviour reported here has so far not been investigated.

With increasing thickness, desorption processes for ASW and ice I$sd$ are shown to be separated by 10–15 K. The QCM data quantifies the H$_2$O loss during crystallisation to be ~4% for a 100 μm thick film heated at ~1 K min$^{-1}$. This means that upon crystallisation of ASW, H$_2$O-ice in an astrophysical environment may only lose a small amount of mass during crystallisation, but this will of course be dependent on the heating rate.

As ASW crystallises, ice I$sd$ forms and not ice I$c$ as commonly stated. The conversion from ASW to ice I$sd$ and later to ice I$h$ are all irreversible upon heating. However, amorphisation of ice I through energetic processing such as proton, ion, and molecular bombardment can still occur (Baratta et al. 1991; Moore and Hudson 1992; Leto and Baratta 2003; Rosu-Finsen and McCoustra 2018). Aside from direct physisorption of ice I$h$ (believed to be a minor part of ice build-up), the formation of ice I$h$ from ice I$sd$ occurs 10-15 K before full desorption for the ≥10 μm thick films. Therefore, it is possible that the main ice I polymorph found in the interstellar medium is ice I$sd$ existing over a 50–60 K range. The spectral differences between ice I$sd$, I$c$ and I$h$ have previously been investigated through Raman spectroscopy, and a red-shift in the O-H stretch has been noted as ice I$sd$ is heated to ice I$h$ which could be of interest to the astronomical community (Carr, Shephard, and Salzmann 2014; del Rosso et al. 2020).

Full ice I$h$ desorption is not observed until 220 K for the 100 μm thick film studied here. Any guest species still trapped within an ice I matrix following crystallisation would only co-desorb with ice I. As full ice I desorption is first reached near 220 K, this gives thick $H_2O$ films the potential to act as a molecular sink retaining species at increasing temperatures that would otherwise have been released into the gas-phase.

## 5 Conclusions

The peak of ASW desorption is found near 150 K with a desorption energy of 64.2 kJ mol$^{-1}$ and a pre-exponential factor $9\times10^{17}$ s$^{-1}$ where ~4% of the $H_2O$ molecules desorb for the 100 μm film. This desorption energy is smaller than that of ice I due to the differences in free energy and vapour pressure.

Following crystallisation, the X-ray diffraction patterns clearly show the formation of ice I$sd$ and not the commonly mentioned ice I$c$. Coupled with QCM and TPD, the X-ray diffraction patterns show that the two-stage desorption in Figure 1(c) and the drop in partial pressures in Figure 3(a), relate to the desorption of ice I$sd$ followed by ice I$h$ at higher temperatures. There is a 50–60 K temperature range between ASW desorption and the ice I$sd$→ice I$h$ transition meaning that a significant amount of crystalline ice I in the astrophysical environment is expected to be ice I$sd$.

Ice I$sd$ desorption starts near 170 K. The desorption kinetics of the 53 nm film in this work matches those reported for thin films studied in ultrahigh vacuum conditions with a desorption energy near 50 kJ mol$^{-1}$ and pre-exponential factor near $10^{30}$ molecules cm$^{-2}$ s$^{-1}$. However, with increasing amounts of $H_2O$, the desorption energy increases rapidly up to about 64–67 kJ mol$^{-1}$ where a plateau is reached as shown in Figure 4(b). The large increase in desorption energy is due to increase the film thickness where the above-layered $H_2O$ molecules act as a physical barrier for the desorbing molecules. Depending on the temperature, thermally activated but physically trapped molecules can re-adsorb forming ice I$h$. As such, the $H_2O$-vacuum-interfacing molecules only desorb once the heat has been transmitted through the bulk of the ice. Eventually the entire film desorbs leading to the apparent desorption energies stated in this work. As $H_2O$ is known to be a good thermal insulator, this transfer takes time where delays in desorption of reaction products or guest species have previously been mentioned for $H_2O$ films (Gadallah et al. 2017; Jiménez-Escobar et al. 2022).

When modelling the effect of radiative heating on the desorption of thick icy layers (Rubanenko and Aharonson 2017; Davidsson and Hosseini 2021), the parameters extracted from desorption experiments are crucial. As radiative heating affects the ice-vacuum interface, the desorption parameters in the literature are ideal. However, in the event of grain core heating or exothermic reactions releasing energy in a $H_2O$ matrix, the energy typically thought to lead to $H_2O$ desorption can be revised to consider film thickness. As $H_2O$ thicknesses vary, the desorption energy increases by up to 30% compared to what is typically reported in the literature meaning that models need to include a desorption energy relevant to the type of heating and film thickness.

## Acknowledgements

We thank M. Vickers for help with the X-ray diffraction measurements and J. K. Cockcroft for access to the Cryojet. This project has received funding from the European Research Council (ERC) under the European Union's Horizon 2020 research and innovation programme (Grant Agreement No. 725271).